\documentclass[aps,twocolumn,reprint,prl,preprintnumbers,superscriptaddress]{revtex4-1}

\usepackage{amsmath,amssymb,amsthm,tikz,tikz-cd,multirow,bm,graphicx,float,array,rotfloat,pgf,soul,bm,xcolor,comment}
\pdfoutput=1

\usepackage[all]{xy}
\usepackage[shortlabels]{enumitem}
\usepackage[normalem]{ulem}
\usepackage[hidelinks]{hyperref}
\usepackage[vcentermath]{youngtab}
\usepackage[boxsize=.8em]{ytableau}

\usetikzlibrary{arrows,arrows.meta,shapes,shapes.misc,patterns,decorations.pathmorphing,decorations.pathreplacing,positioning,chains,fit,bending,decorations.markings,intersections}
\tikzset{
    gauge/.style={rounded rectangle, draw=black!100, thick, minimum size=5mm},  
    gaugeD/.style={rounded rectangle, draw=black!100,double,thick,minimum size=5mm}, 
    empty/.style={rounded rectangle, draw=white!100, thick, minimum size=5mm}, 
    flavor/.style={rectangle, draw=black!100, thick, minimum size=5mm},
    arc arrow/.style args={%
    to pos #1 with length #2}{
    decoration={
        markings,
         mark=at position 0 with {\pgfextra{%
         \pgfmathsetmacro{\tmpArrowTime}{#2/(\pgfdecoratedpathlength)}
         \xdef\tmpArrowTime{\tmpArrowTime}}},
        mark=at position {#1-\tmpArrowTime} with {\coordinate(@1);},
        mark=at position {#1-2*\tmpArrowTime/3} with {\coordinate(@2);},
        mark=at position {#1-\tmpArrowTime/3} with {\coordinate(@3);},
        mark=at position {#1} with {\coordinate(@4);
        \draw[-{Stealth[length=#2,bend]}]       
        (@1) .. controls (@2) and (@3) .. (@4);},
        },
     postaction=decorate,
     },arr/.style={arc arrow=to pos #1 with length 2.3mm}
}
\theoremstyle{plain}
\newtheorem*{thm*}{Theorem}

\def\CN{\mathcal{N}}

\def\CD{\mathcal{D}}
\def\ft{\mathfrak{t}}
\def\tr{\operatorname{Tr}}
\begin{document}
\title{Emergent \texorpdfstring{$\CN=4$}{N=4} supersymmetry from \texorpdfstring{$\CN=1$}{N=1}}

\vspace*{-3cm} 
\begin{flushright}
{\tt CALT-TH-2023-005}\\
{\tt DESY-23-021}\\
\end{flushright}
\author{Monica Jinwoo Kang}
\email[\texttt{monica@caltech.edu}]{}
\affiliation{Walter Burke Institute for Theoretical Physics, California Institute of Technology, Pasadena, CA 91125, U.S.A.}

\author{Craig Lawrie}
\email[\texttt{craig.lawrie1729@gmail.com}]{}
\affiliation{Deutsches Elektronen-Synchrotron DESY, Notkestr.~85, 22607 Hamburg, Germany}

\author{Ki-Hong Lee}
\email[\texttt{khlee11812@gmail.com}]{}
\affiliation{Department of Physics, Korea Advanced Institute of Science and Technology, Daejeon 34141, Republic of Korea}

\author{Jaewon Song}
\email[\texttt{jaewon.song@kaist.ac.kr}]{}
\affiliation{Department of Physics, Korea Advanced Institute of Science and Technology, Daejeon 34141, Republic of Korea}

\begin{abstract}
\noindent 
We discover a four-dimensional $\CN=1$ supersymmetric field theory that is dual to the $\CN=4$ super Yang--Mills theory with gauge group $SU(2n+1)$ for each $n$. The dual theory is constructed through the diagonal gauging of the $SU(2n+1)$ flavor symmetry of three copies of a strongly-coupled superconformal field theory (SCFT) of Argyres--Douglas type.
We find that this theory flows in the infrared to a strongly-coupled $\mathcal{N}=1$ SCFT that lies on the same conformal manifold as $\mathcal{N}=4$ super Yang--Mills with gauge group $SU(2n+1)$. Our construction provides a hint on why certain $\CN=1, 2$ SCFTs have identical central charges ($a=c$).
\end{abstract}

\maketitle

%
%
%

\section{Introduction}

Symmetry is the fundamental organizing principle used to characterize physical systems. Symmetry can be spontaneously broken or emergent along renormalization group (RG) flows. In particular, supersymmetry is a powerful symmetry, which exchanges bosons and fermions in spacetime, that is traditionally regarded as a high-energy symmetry in the ultraviolet (UV). 
Alternatively, supersymmetry can arise as an emergent symmetry in the infrared (IR). This idea has been studied in various situations.
In two dimensions, supersymmetry has been shown to emerge in the dilute Ising model at the tricritical point \cite{Friedan:1984rv}; this feature of emergent supersymmetry has been extended to quantum critical points of higher-dimensional lattice models \cite{Lee:2006if}. In four dimensions, it has been suggested that supersymmetric Yang--Mills (SYM) theory can arise from strong-coupling dynamics in the low energy limit of a non-supersymmetric gauge theory \cite{Kaplan:1983sk}. Also, it has been suggested that  supersymmetry can emerge at the edges of a topological superconductor, that can be potentially realized experimentally \cite{Grover:2013rc}. 

Instead of starting without any supersymmetry, one can begin with minimal supersymmetry and flow to an enhanced supersymmetry in the IR \cite{Gadde:2015xta, Maruyoshi:2016tqk, Razamat:2019vfd, Zafrir:2019hps, Razamat:2020gcc, Zafrir:2020epd}. This is not only an interesting phenomenon by itself, but also provides a powerful tool to analyze non-perturbative dynamics of the IR fixed point, which often has no Lagrangian description with the full extended supersymmetry manifest. In this sense, supersymmetry enhancement can be thought of as an IR duality. 
IR duality refers to the phenomenon where there are two distinct UV theories that become identical in the IR after the RG flow. A well-known example is Seiberg duality \cite{Seiberg:1994pq}, under which the fundamental fields in a gauge theory are mapped to composite fields of a different gauge theory, and vice versa. 



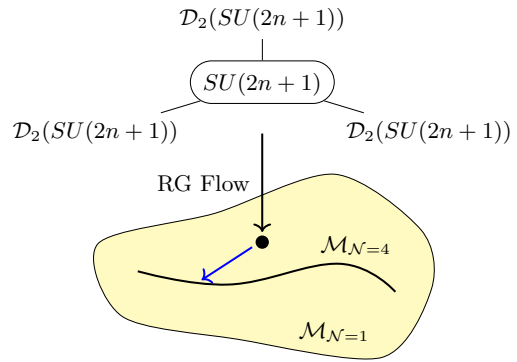
\begin{figure}[h]
    \centering
    \scalebox{.96}{
    \begin{tikzpicture}[declare function={rr=2*(1+0.2*sin(3*\t-10)+0.1*rnd);},xscale=.7,yscale=.35]
    \draw[scale=1.6,rotate=30,fill=yellow!30] plot[smooth cycle,variable=\t,samples at={0,45,...,315},xshift=-6mm,yshift=-40mm] (\t:rr);
    \node[anchor=south west,yshift=4mm](At) at (0.2,2) {$\mathcal{D}_{2}(SU(2n+1))$};
    \node[anchor=south west,yshift=3mm](A1) at (-3.1,-2.2) {$\mathcal{D}_{2}(SU(2n+1))$};
    \node[anchor=south west, draw,rounded rectangle, inner sep=3pt, minimum size=5mm, text height=3mm,yshift=2mm](A2) at (1,0) {$SU(2n+1)$};
    \node[anchor=south west,yshift=3mm](A3) at (3.5,-2.2) {$\mathcal{D}_{2}(SU(2n+1))$};
    \draw (A1)--(A2)--(A3);
    \draw (At)--(A2);
    \draw[->,thick] (2,-.6)--(2,-4.5) node[midway]{RG Flow$\qquad\qquad\quad$};
    \node[circle, scale=.6, fill=black] at (2,-4.9) {}; \draw[thick,xshift=50mm,yshift=-69mm,rotate=220,scale=1.3] plot [smooth, tension=.7] coordinates { (.2,-.2) (0.3,-1.4) (2,-2) (2.8,-3.2)};
    \draw[->,thick,blue] (1.8,-5.1)--(.8,-6.4) node[midway] {};
    \node[anchor=south west,xshift=23mm,yshift=-20mm] at (-.3,0) {$\mathcal{M}_{\mathcal{N}=4}$};
    \node[anchor=south west,xshift=20mm,yshift=-32mm] at (-.3,0) {$\mathcal{M}_{\mathcal{N}=1}$};
    \end{tikzpicture}
    }
    \caption{The 4d $\CN=1$ gauge theory flows to a point in the $\CN=1$ preserving conformal manifold ($\mathcal{M}_{\mathcal{N}=1}$) of the $SU(2n+1)$ $\CN=4$ SYM theory. We emphasize that the flow from the gauge theory into the infrared generically leads to an $\mathcal{N}=1$ SCFT; an exactly-marginal deformation then moves to a point on the $\mathcal{N}=4$ preserving conformal manifold ($\mathcal{M}_{\mathcal{N}=4}$), and thus an enhancement to $\mathcal{N}=4$ supersymmetry.}
    \label{fig:RGflow}
\end{figure}

In this Letter, we find a novel dual description to $\CN=4$ SYM that exhibits emergent maximal supersymmetry. The dual theory involves non-Lagrangian four-dimensional $\CN=2$ superconformal theories (SCFTs) coupled to an $\CN=1$ gauge multiplet. This theory then flows in the IR to an SCFT, and we argue that the fixed point lies on the $\CN=1$ preserving conformal manifold \cite{Leigh:1995ep} of $\CN=4$ SYM. This is depicted in Figure \ref{fig:RGflow}.

The dual description we present is interesting in many ways. First, we find that the $\CN=1$ supersymmetry in the UV gets enhanced to $\CN=4$ in the IR, which is a larger enhancement compared to the $\CN=2$ enhancement discussed in \cite{Gadde:2015xta, Maruyoshi:2016tqk, Razamat:2019vfd, Zafrir:2019hps, Razamat:2020gcc, Zafrir:2020epd}. Second, the $\CN=1$ dual theory is given in terms of non-Lagrangian SCFTs of Argyres--Douglas type \cite{Argyres:1995jj, Xie:2012hs, Cecotti:2012jx, Cecotti:2013lda}, however it flows to a theory with a Lagrangian description; this provides a hint on the earlier observations that the relevant Argyres--Douglas theory in this construction ($\mathcal{D}_2(G)$) behaves similarly to free fields \cite{Xie:2016evu, Song:2017oew, Buican:2017rya}. This generalizes the previous $\mathcal{N}=1$ duals to $\mathcal{N}=4$ SYM with gauge group $SU(2)$, where the $\mathcal{N}=1$ theory is Lagrangian \cite{Intriligator:1995id}.
Finally, the $\CN=4$ enhancement of supersymmetry provides an explanation of recently explored $a=c$ theories with $\CN=1, 2$ supersymmetry \cite{Kang:2021lic, Kang:2021ccs,Kang:2022zsl,Kang:2022vab,LANDSCAPE}. 

\section{The dual theory}

To construct the non-Lagrangian $\mathcal{N}=1$ duals to $\mathcal{N}=4$ SYM, we introduce some 4d $\mathcal{N}=2$ SCFTs as ``building blocks.'' We consider 
the theories known as $\mathcal{D}_p(G)$, where $p \geq 2$ is an integer and $G$ is an ADE Lie algebra \cite{Cecotti:2012jx,Xie:2012hs,Cecotti:2013lda}. This theory contains (at least) $G$ as its flavor symmetry. Such SCFTs can be constructed 
from the class $\mathcal{S}$ \cite{Gaiotto:2009we,Gaiotto:2009hg, Xie:2012hs} perspective, that is by compactifying the 6d $\CN=(2, 0)$ theory of type $G$ on a sphere with certain punctures, one labelled by $p$.
These $\mathcal{N}=2$ SCFTs are generically inherently strongly-coupled as they possess Coulomb branch 
operators with fractional conformal dimensions; such SCFTs are known as 
Argyres--Douglas theories following the seminal example in \cite{Argyres:1995jj}. 

In this Letter, we consider the diagonal gauging of three copies of $\mathcal{D}_2(SU(2n+1))$ via an $\CN=1$ gauge multiplet, as depicted at the top of Figure \ref{fig:RGflow}. 
The one-loop $\beta$-function of the gauge coupling at the central node is given by the anomaly between the R-symmetry and the gauge current $\beta_g \sim - \mathrm{Tr} RGG \big|_{UV} \sim - \frac{3}{2}(2n+1)$, which is negative and thus the theory is asymptotically-free \cite{Kang:2021ccs}. In the infrared, we claim that we obtain an interacting 4d $\mathcal{N}=1$ SCFT that is on the conformal manifold of $\CN=4$ super Yang--Mills, as we now explain.

Under the $\mathcal{N}=1$ gauging the $SU(2)_R^{\mathcal{N}=2} \times U(1)_R^{\mathcal{N}=2}$ R-symmetry of each of the $\mathcal{D}_2(SU(2n+1))$ factors decomposes into a $U(1)_R$ and a $U(1)_F$, with generators
\begin{equation}
    R_0 = \frac{1}{3}r + \frac{4}{3}I_3 \,, \qquad F = -r + 2I_3 \,,
\end{equation}
where $r$ is the generator of $U(1)_R^{\mathcal{N}=2}$ and $I_3$ is the generator of the Cartan of the $SU(2)_R^{\mathcal{N}=2}$. Upon gauging, the putative $U(1)_{R_0}\times U(1)_F^3$ global symmetry has an ABJ anomaly, and it is broken to the anomaly-free $U(1)_R \times U(1)_{\mathcal{F}}^2$, generated by certain linear combinations of $R_0$ and the $F_i$, where $i=1,2,3$ runs over the $\mathcal{D}_2(SU(2n+1))$ factors. The flavor symmetries are generated by $\mathcal{F}_1 \equiv F_2 - F_1$ and $\mathcal{F}_2\equiv F_3 - F_2$.
At the IR fixed point, the superconformal R-symmetry is 
\begin{equation}\label{eqn:RIR}
    R = R_0 + \sum_{i=1}^3 \varepsilon_i F_i \,,
\end{equation}
where the $\varepsilon_i$ are fixed by the anomaly-free condition of the R-charge and $a$-maximization \cite{Intriligator:2003jj}. The $a$-maximization procedure is carried out in \cite{Kang:2021ccs} and we repeat this in the following section. 

Furthermore in \cite{Kang:2021ccs, Kang:2022vab}, the authors confirm that all the low-lying operators satisfy the unitarity constraint. Thus the non-Lagrangian $\mathcal{N}=1$ theory depicted in Figure \ref{fig:RGflow} flows in the IR to an interacting SCFT. We provide evidence that this theory is conformally dual to $\CN=4$ SYM in the following sections. 

\section{Matching anomalies}

The anomaly polynomial of the putative dual four-dimensional $\mathcal{N}=1$ SCFT takes the form
\begin{equation}\label{eqn:APgeneric}
  \begin{aligned}
    I_6 &= \frac{1}{6}k_{RRR}c_1(R)^3 + \sum_{\alpha=1}^2\frac{1}{6}k_{RR\mathcal{F}_\alpha}c_1(R)^2c_1(\mathcal{F}_\alpha) \\&\quad + \sum_{\alpha,\beta = 1}^{2}\frac{1}{6}k_{R\mathcal{F}_\alpha\mathcal{F}_\beta}c_1(R)c_1(\mathcal{F}_\alpha)c_1(\mathcal{F}_\beta) \\&\quad+ \sum_{\alpha,\beta,\gamma=1}^2\frac{1}{6}k_{\mathcal{F}_\alpha\mathcal{F}_\beta\mathcal{F}_\gamma}c_1(\mathcal{F}_\alpha)c_1(\mathcal{F}_\beta)c_1(\mathcal{F}_\gamma) \\&\quad - \frac{1}{24}k_Rc_1(R)p_1(T) - \sum_{\alpha=1}^2 \frac{1}{24}k_{\mathcal{F}_\alpha}c_1(\mathcal{F}_\alpha)p_1(T) \,,
  \end{aligned}
\end{equation}
where $c_1(R)$ is the first Chern class of the superconformal R-symmetry bundle, $p_1(T)$ is the first Pontryagin class of the tangent bundle to the 4d spacetime, and $c_1(\mathcal{F}_\alpha)$ is the first Chern class of the bundles associated to each $U(1)$ flavor symmetry factor. 
 
In a supersymmetric theory, the conformal anomalies or central charges are related to the anomaly coefficients of the R-symmetry \cite{Anselmi:1997am}
\begin{equation} \label{eqn:ac}
    a = \frac{3}{32}(3k_{RRR} - k_R) \,, \quad c = \frac{1}{32}(9k_{RRR} - 5k_R) \,.
\end{equation}
Utilizing equation \eqref{eqn:RIR}, the anomaly coefficients can be written in terms of the anomaly coefficients of $R_0$ and $F_i$. Since $F_i$ acts only on the $i^\text{th}$ $\mathcal{D}_2(SU(2n+1))$, the mixed anomaly coefficients $k_{R_0F_iF_j}$, $k_{F_i^2F_j}$, and $k_{F_iF_jF_k}$ vanish and each anomaly coefficient is a linear combination of those of  each $\mathcal{D}_2(SU(2n+1))$:
\begin{equation}\label{eqn:anom_D2G}
\begin{gathered}
    k_{rrr}=k_{r}=48(a-c)\,,\quad k_{rI_3I_3}=4a-2c\,,\\
    a=\frac{7}{96}d\,,\quad c=\frac{1}{12}d\,,
\end{gathered}
\end{equation}
together with the contribution of the vector multiplet. Here, $d=4n(n+1)$ denotes the dimension of $SU(2n+1)$. 

Then we can compute the trial $a$-function for the trial R-symmetry in equation \eqref{eqn:RIR} using equations \eqref{eqn:ac} and \eqref{eqn:anom_D2G}, which is given as 
\begin{align}
a(\varepsilon_1,\varepsilon_2,\varepsilon_3)=\frac{d}{32}\left(13-9\sum_{i=1}^3\varepsilon_i^2(\varepsilon_i+2)\right) \ . 
\end{align}
Combined with the anomaly-free condition for the R-symmetry $6-\sum_{i=1}^3 \left(1-3\varepsilon_i\right) =0$, $a$-maximization fixes the mixing coefficients: 
\begin{equation}\label{eqn:epsilon}
    \varepsilon := \varepsilon_1 = \varepsilon_2 = \varepsilon_3 = - \frac{1}{3} \,.
\end{equation}
Then, it is straightforward to determine the 't Hooft anomaly coefficients in equation \eqref{eqn:APgeneric},
\begin{align}
\begin{gathered}
    k_{RRR}=\frac{8d}{9}\,, \quad k_{R\mathcal{F}_{\alpha}^2}=-\frac{2d}{3}\,,\quad k_{R\mathcal{F}_1\mathcal{F}_2}=\frac{d}{3} \ , \\
    k_{\mathcal{F}_1^2\mathcal{F}_2}=-k_{\mathcal{F}_1\mathcal{F}_2^2}=d\,,\quad a=c=\frac{1}{4}d\,,
\end{gathered}
\end{align}
while the others are zero. 
These match exactly with the anomaly coefficients of the $\mathcal{N}=4$ $SU(2n+1)$ gauge theory, where each $\mathcal{F}_{\alpha}$ corresponds to the Cartan of the, $SU(3)$ flavor symmetry rotating the adjoint chirals. Since the 't Hooft anomalies are invariant on a conformal manifold, it supports the claim that the theory in Figure \ref{fig:RGflow} flows in the infrared to the same conformal manifold as that of the $\CN=4$ SYM theory. 

\section{Matching chiral operators}

Let us now compare the gauge invariant operator spectrum of the dual theories. The single-trace chiral operators of $\CN=4$ SYM are given in the form of 
$ \operatorname{Tr} \phi_{i_1} \phi_{i_2} \cdots \phi_{i_k} $ with $i_1, \ldots , i_k \in \{ 1, 2, 3 \} $, where $\phi _{1, 2, 3}$ are adjoint chiral multiplets. 
We find that these operators are mapped to 
\begin{equation}
    \operatorname{Tr} \phi_{i_1} \phi_{i_2} \cdots \phi_{i_k} \,\,\rightarrow \,\,
    \operatorname{Tr} \mu_{i_1} \mu_{i_2} \cdots \mu_{i_k} \,,
\end{equation}
where $\mu_{1, 2, 3}$ is the moment map operator in each $\mathcal{D}_2(SU(2n+1))$ theory that we gauge. This map follows since the scaling dimension of the moment map operators (carrying R-charges $I_3 = 1, r=0$) becomes 
$ \Delta_{\mathrm{IR}}(\mu) = \frac{3}{2} R = \frac{3}{2} \left( \frac{4}{3}+2 \varepsilon \right) = 1$
in the IR under the RG flow, which can be computed using equations \eqref{eqn:RIR} and \eqref{eqn:epsilon}. 

Among the gauge invariant chiral operators, $\operatorname{Tr}(\phi_i)^k$ (Casimir operators) are mapped to the Coulomb branch operators of the $\mathcal{D}_2(SU(2n+1))$ theory as we explain below. 
Each $\mathcal{D}_2(SU(2n+1))$ theory contains Coulomb branch operators, whose conformal dimensions are
\begin{equation}\label{eqn:D2CB}
    \Delta_\text{CB} = \left\{ \frac{3}{2}, \frac{5}{2}, \cdots, \frac{2n+1}{2} \right\} \,.
\end{equation}
Each $\mathcal{N}=2$ multiplet containing a Coulomb branch operator $u$ 
has two $\CN=1$ chiral multiplets (with scalar primaries), whose scalar components are $u$ and its $\CN=2$ descendant that we call as $Q^2u$, generated by the $\mathcal{N}=2$ supercharge $Q$ (that is broken after $\CN=1$ gauging). 

Upon flowing the gauge theory in Figure \ref{fig:RGflow} into the IR, the conformal dimensions of $u$ are shifted as
\begin{equation}
    \Delta_\text{IR}(u) = \left(1-3\varepsilon\right)\Delta_\text{UV}(u) \,,
\end{equation}
and the conformal dimensions of the $Q^2u$ operators become
\begin{equation}
    \Delta_\text{IR}(Q^2u)= 1 + 6 \varepsilon +\left(1 -3\varepsilon\right)\Delta_\text{UV}(u) \,.
\end{equation}
Upon plugging in $\varepsilon = -1/3$, the Coulomb branch operators ($u, Q^2 u$) belonging to each one of the three Argyres--Douglas theories become chiral operators of conformal dimensions 
$ \Delta = \left\{ 2, 3, \cdots, 2n + 1 \right\} $. 
They are mapped to the operators $\operatorname{Tr} (\phi_i)^k$ in the $\CN=4$ SYM theory. 

An important property of the $\mathcal{D}_2(SU(2n+1))$ theory is that the operator $\mu$ enjoys chiral ring relations  
\begin{equation}\label{eqn:crr}
    \mu^2 \big{|}_{\textrm{adj}} = 0 \,, \quad \mathrm{Tr} \mu^k = 0 \,,
\end{equation}
for each $k$. The first relation says that the adjoint part of the square of moment map operator is vanishing, which can be obtained from the Schur index \cite{Xie:2016evu, Song:2017oew, Agarwal:2018zqi}. The second relation arises because the Higgs branch of this theory is given by a nilpotent orbit. The second chiral ring relation in equation \eqref{eqn:crr} removes the Casimir operators in the spectrum that would be superfluous. 

Our dual theory has five marginal operators: three of them come from each of the Coulomb branch operators of dimension $3/2$ (in the UV), and two of the form $\operatorname{Tr}\mu_1 \mu_2 \mu_3$ and $\operatorname{Tr}\mu_1 \mu_3 \mu_2$. 
Other candidate marginal combinations of the $\mu_i$ are removed via the chiral ring relation in equation \eqref{eqn:crr}.
Two of these operators are marginally irrelevant as they break the $U(1)^2$ flavor symmetry. Each broken flavor symmetry generator combines with a marginal operator, forms a long (non-BPS) multiplet, and becomes irrelevant \cite{Green:2010da}. The other three operators are exactly marginal, their vacuum expectation values parametrize a complex manifold, and this is precisely the three-complex-dimensional conformal manifold of the infrared theory. 

Now we compare with the conformal manifold of $\CN=4$ SYM.
In $\CN=4$ SYM, there are eleven marginal operators of the form $\operatorname{Tr}\phi_i \phi_j \phi_k$. Among the eleven, eight of them recombine with the generators of the $SU(3)$ flavor symmetry that is broken at the generic point of the conformal manifold, and thus are marginally irrelevant \footnote{There is also a gauge coupling, which breaks a $U(1)$ symmetry via an ABJ anomaly \cite{Leigh:1995ep}. Therefore the counting of the dimension still goes through.}. 
Hence, the conformal manifold of 4d $\mathcal{N}=4$ super Yang--Mills has complex-dimension three, matching that of the $\mathcal{N}=1$ dual. When we move to the $U(1)^2$ preserving sublocus in the conformal manifold, the off-diagonal generators of the $SU(3)$ current combine with marginal operators to become long multiplets and become irrelevant. This removes six out of eleven marginal operators, leaving only five, matching that of the dual theory. 

\section{Superconformal index}

In this section, we compute the superconformal index to test the duality beyond the chiral ring. 
The superconformal index \cite{Kinney:2005ej,Romelsberger:2005eg} is defined as
\begin{align}\label{eqn:N=1idx}
I=\operatorname{Tr}(-1)^Ft^{3(R+2j_2)}y^{2j_1}\prod_{i}v_i^{f_i}\,,
\end{align}
where the trace is taken over the states satisfying the condition on the conformal dimension: $\Delta=\frac{3}{2}R+2j_2$, and where $j_{1,2}$ are the Lorentz spins, $R$ is the R-charge, and the flavor charges are denoted by $f_i$. 
The index is invariant under the RG flow, enabling us to compute the BPS spectrum in the IR from the UV description. 

From the $\CN=1$ gauge theory description of the $\mathcal{D}_2(SU(3))$ theory \cite{Maruyoshi:2016tqk, Maruyoshi:2016aim, Agarwal:2016pjo}, the (reduced) superconformal index of the IR SCFT is computed \cite{Kang:2022vab} as
\begin{align}
\begin{aligned}
&\widehat{I}^{\mathfrak{su}_3} \equiv (1-t^3y)(1-t^3/y)(I^{\mathfrak{su}_3}-1) \\
&\quad = t^4\chi^{\mathfrak{su}_3}_{\mathbf{6}}-t^5\chi^{\mathfrak{su}_2}_{\mathbf{2}}\chi^{\mathfrak{su}_3}_{\mathbf{3}} +t^6 (\chi^{\mathfrak{su}_3}_{\mathbf{10}}-\chi^{\mathfrak{su}_3}_{\bf{8}}+1)\\
&\quad  -t^7\chi^{\mathfrak{su}_2}_{\bf{2}} (\chi^{\mathfrak{su}_3}_{\bf{6}}-\chi^{\mathfrak{su}_3}_{\overline{\bf{3}}}) +t^8\small( \chi^{\mathfrak{su}_3}_{\bf{15'}}-\chi^{\mathfrak{su}_3}_{\bf{15}}+\chi^{\mathfrak{su}_3}_{\overline{\bf{6}}}\\
&\quad
\left.+2\chi^{\mathfrak{su}_3}_{\bf{3}}\right)-t^9\chi^{\mathfrak{su}_2}_{\bf{2}}\left(\chi^{\mathfrak{su}_3}_{\bf{10}}+1\right)+t^{10}\chi^{\mathfrak{su}_2}_{\bf{3}}\chi^{\mathfrak{su}_3}_{\overline{\bf{3}}}\\
&\quad +t^{10} (\chi^{\mathfrak{su}_3}_{\overline{\bf{21}}}-\chi^{\mathfrak{su}_3}_{\overline{\bf{15}}}+2\chi^{\mathfrak{su}_3}_{\bf{6}}-2\chi^{\mathfrak{su}_3}_{\overline{\bf{3}}}) 
+\cdots ,
\end{aligned}
\end{align}
where $\chi^{\mathfrak{su}_2}_{\bf{R}}=\chi^{\mathfrak{su}_2}_{\bf{R}}(y)$ is the character of the representation $\bf{R}$ in Lorentz spin $j_1$. The $U(1)^2$ flavor symmetry enhances to $SU(3)$ at certain points of the conformal manifold, thus each term is written in terms of the character $\chi^{\mathfrak{su}_3}_{\bf{R}}$ of the representation $\bm{R}$ of the enhanced flavor. 
This exactly matches with the superconformal index of $\mathcal{N}=4$ SYM with gauge group $SU(3)$: the $SU(3)$ flavor is a part of $SO(6)$ R-symmetry rotating the three adjoint chiral multiplets.

The full index for the $\mathcal{D}_2(SU(2n+1))$ theory is not available for $n>1$. However, a certain limit of the index is available to test the duality. 
The superconformal index for an $\CN=2$ SCFT can be written as 
\begin{align}
    I_{\CN=2}(p, q, \mathfrak{t}) = \tr (-1)^F p^{j_1 + j_2 + r} q^{j_2 - j_1 + r} \ft^{I_3 - \frac{1}{2} r} \,, 
\end{align}
where we take the trace over the states with $\Delta - 2j_2 - 2I_3 - r/2=0 $. The Schur limit of this index is obtained by taking $q= \ft$. In this limit, the $p$-dependence drops out \cite{Gadde:2011uv}. The Schur limit of the index for $\CD_2(SU(2n+1))$ is available and given by a very succinct formula \cite{Xie:2016evu, Song:2017oew}
\begin{align} \label{eqn:D2Schur}
    I_S^{\mathcal{D}_2(SU(2n+1))} (q; z) = \textrm{PE} \left[ \frac{q}{1-q^2} \chi_{\mathbf{adj}} (z) \right] \ , 
\end{align}
where PE denotes the plethystic exponential. It was found in \cite{Buican:2016hnq} that one can take a special limit of the superconformal index for an $\CN=1$ deformed $\CN=2$ SCFT, such that it agrees with the Schur index.

In our case, when we glue the $\CD_2(SU(2n+1))$ theories via  an $\CN=1$ vector multiplet, the index for the IR SCFT can be computed once we know the $\CN=2$ expression of the index. If the superconformal R-symmetry is given as in equation \eqref{eqn:RIR}, the superconformal index is schematically given as
\begin{align}
    I(p, q) = \int [dz] I_{\text{vec}}(z) \prod_{i=1}^3 I^{\CD_2} (z)\Big|_{\ft \to  (pq)^{\frac{2}{3}+\varepsilon_i}} \,,
\end{align}
where $p$ and $q$ are $t^3y$ and $t^3/y$ in equation \eqref{eqn:N=1idx}, $I_{\text{vec}}(z)$ and $I^{\CD_2}(z)$ denote the index for an $\CN=1$ vector multiplet and the $\mathcal{D}_2(SU(2n+1))$ theory respectively, and $\varepsilon_i = -1/3$ from equation \eqref{eqn:epsilon}. 

We would like to compare this against the index of the $\CN=4$ SYM theory, which is
\begin{align}
    I^{\CN=4}(p, q) = \int [dz] I_{\text{vec}}(z) I_{\text{chi}}(z)^3 \,,
\end{align}
where $I_{\text{chi}}(z)$ denotes the index for the adjoint chiral:
\begin{align} \label{eqn:idx_adjchiral}
    I_{\text{chi}}(z) = \text{PE} \left[ \frac{(pq)^{1/3} - (pq)^{2/3}}{(1-p)(1-q)} \chi_{\textbf{adj}}(z) \right] \,. 
\end{align}
Now, we take the limit $q = \ft = (pq)^{\frac{1}{3}}$ so that $p \to q^2$. In this limit, the index $I^{\CD_2}(z)$ reduces to the Schur index in equation \eqref{eqn:D2Schur} and the free chiral multiplet index in equation \eqref{eqn:idx_adjchiral} becomes identical to the Schur index of the $\CD_2(SU(2n+1))$ theory as in equation \eqref{eqn:D2Schur}. Therefore in this limit, the index of our gauge theory, for any $n \geq 1$, agrees with that of $\CN=4$ SYM. 

\section{Comments on the \texorpdfstring{\boldmath{$G \neq SU(2n+1)$}}{G!=SU(2n+1)} case}

We have thus far proposed a dual theory to the $\CN=4$ SYM with gauge group $G=SU(2n+1)$. It is natural to ask if there exist similar dualities for other gauge groups. While we were not able to find a direct analog of such a dual description, we find that there indeed exist RG flows that connect the $\CN=1, 2$ SCFTs with $a=c$ in \cite{Kang:2021lic, Kang:2021ccs} to $\CN=4$ SYM upon a series of relevant deformations. 

A family of $\CN=1$ SCFTs with $a=c$ can be constructed out of a diagonal gauging of $\mathcal{D}_{p_i\geq 2}(G)$ theories whose flavor symmetry is precisely $G\in \text{ADE}$. Then, for a choice of $G$, a positive integer $p_i$ can be chosen accordingly. The $\CD_{p}(G)$ theory has a Coulomb branch operator $u$ of scaling dimension $\Delta(u) = \frac{p+1}{p}$. By taking a linear deformation $W=u$, the $\CD_p(G)$ theory flows to a theory of $|G|$ free chiral multiplets \cite{Bolognesi:2015wta,Xie:2021omd}. 
Hence, starting with a theory of three $\mathcal{D}_{p_i}(G)$ diagonally gauged, we can perform a series of such linear deformations on each block to land on a gauge theory with three adjoint chiral multiplets, which is $\CN=4$ SYM (modulo marginal deformations). When considering more than three $\mathcal{D}_{p_i}(G)$, we can simply integrate out the adjoint chiral multiplet after the linear deformation. In this way, we can connect (not all but) many of the $a=c$ theories constructed via gluing $\CD_p(G)$ theories to $\CN=4$ SYM via RG flow \footnote{When more than three $\mathcal{D}_{p_i}(G)$ theories are glued, it flows to $\mathcal{N}=4$ SYM only for restricted choices of $p_i$; most become IR-free upon linear deformation \cite{LANDSCAPE}.}. 

It is found in \cite{Kang:2021ccs, LANDSCAPE} that $a=c$ is preserved under any $\mathcal{N}=1$ preserving superpotential deformation, via a relevant operator, of the $a=c$ SCFTs constructed by diagonal gaugings of $\mathcal{D}_{p_i\geq 2}(G)$ theories in \cite{Kang:2021lic,Kang:2021ccs}. This works as long as there is no accidental symmetry. When a (gauged) $\CD_p(G)$ theory gets deformed and flows to free chiral multiplets, the original $U(1)_F$ symmetry is broken but a new symmetry that acts on the free fields emerges. Since the R-symmetry can potentially mix with this new symmetry, it is not guaranteed to preserve $a=c$ along the flow. 
Surprisingly, we nevertheless find in our constructions of $a=c$ theories that $a=c$ is  preserved under this flow to $\mathcal{N}=4$ SYM.

\section{Discussion}

In this Letter, we have constructed a non-Lagrangian $\CN=1$ gauge theory (via diagonally gauging three $\mathcal{D}_2(SU(2n+1))$) that flows to a point on the conformal manifold of the $\CN=4$ SYM with gauge group $SU(2n+1)$, and thereby exhibits maximal supersymmetry enhancement. We verified this by matching anomalies, operators, and superconformal indices. We generalize this to find that many of the $a=c$ theories constructed in \cite{Kang:2021ccs, Kang:2021lic} are connected to $\CN=4$ SYM under RG flows; this sheds light on the origin of $a=c$ without any obvious protection from symmetry. 

Relatedly, a host of class $\mathcal{S}$ theories with $a=c$ has recently appeared in \cite{Kang:2022zsl}; it would be interesting to explore if this orthogonal construction also gives rise to theories that possess deformations to $\mathcal{N}=4$ SYM.

We expect that the $\mathcal{N}=1$ dual theories can be investigated further utilizing the $\mathcal{N}=1$ class $\mathcal{S}$ framework \cite{Bah:2012dg}. Our theory can be realized in class $\mathcal{S}$ by gluing the class $\mathcal{S}$ description of two copies of $\CD_2(SU(2n+1))$ and that of a single $\CD_2(SU(2n+1))$. The former can be obtained via $A_{2n}$ theory on a sphere with three regular punctures, labeled by partitions $[2n]_t, [2n]_t, [1^{2n+1}]$ respectively \cite{Beem:2020pry}. The latter can be realized as a sphere with an irregular puncture and a full regular puncture \cite{Xie:2012hs}. We glue them along the full puncture (partition $[1^{2n+1}]$) with an $\CN=1$ vector to form a sphere with two regular punctures and an irregular puncture. Having this class $\mathcal{S}$ description enables various interesting applications such as computing the index of the $\mathcal{N}=1$ dual theories for $n > 1$ \cite{Beem:2012yn}.

The aforementioned class $\mathcal{S}$ description paves the way to a holographic dual of our $\mathcal{N}=1$ theory in the UV. The putative holographic dual description of $\mathcal{N}=1, 2$ SCFTs with $a=c$ \cite{Kang:2021ccs, Kang:2021lic} should have vanishing four-derivative term $R^{\mu\nu\rho\sigma}R_{\mu\nu\rho\sigma}$ (whose coefficient is proportional to $c-a$ \cite{Anselmi:1998zb}) in the supergravity action. Our novel RG flow suggests that there is a domain wall solution interpolating the $\mathcal{N}=1$ UV theory and the IR $\mathcal{N}=4$ SYM, which admits ``miraculous cancellations,'' and therefore sheds light on this seemingly fine-tuned coefficient. 


In addition, the simplicity of $\mathcal{N}=4$ SYM may illuminate the non-Lagrangian $\mathcal{D}_p(G)$ theories. As $\CN=4$ SYM is a Lagrangian theory, this hints that it may be feasible to extract quantitative features of $\mathcal{D}_p(G)$ from this duality. 

\vspace{0.25cm}
\begin{acknowledgements}

We thank Jacques Distler for helpful discussions. 
C.L.~and J.S.~thank the California Institute of Technology for hospitality where this work was conceived. M.J.K.~is supported by Sherman Fairchild Postdoctoral Fellowship and the U.S.~Department of Energy, Office of Science, Office of High Energy Physics, under Award Number DE-SC0011632. C.L.~acknowledges support from DESY (Hamburg, Germany), a member of the Helmholtz Association HGF. K.H.L.~and J.S.~are supported by the National Research Foundation of Korea (NRF) grant 2020R1C1C1007591, RS-2023-00208602, and also by POSCO Science Fellowship of POSCO TJ Park Foundation.

\end{acknowledgements}


\bibliography{references}

\begin{thebibliography}{45}%
\makeatletter
\providecommand \@ifxundefined [1]{%
 \@ifx{#1\undefined}
}%
\providecommand \@ifnum [1]{%
 \ifnum #1\expandafter \@firstoftwo
 \else \expandafter \@secondoftwo
 \fi
}%
\providecommand \@ifx [1]{%
 \ifx #1\expandafter \@firstoftwo
 \else \expandafter \@secondoftwo
 \fi
}%
\providecommand \natexlab [1]{#1}%
\providecommand \enquote  [1]{``#1''}%
\providecommand \bibnamefont  [1]{#1}%
\providecommand \bibfnamefont [1]{#1}%
\providecommand \citenamefont [1]{#1}%
\providecommand \href@noop [0]{\@secondoftwo}%
\providecommand \href [0]{\begingroup \@sanitize@url \@href}%
\providecommand \@href[1]{\@@startlink{#1}\@@href}%
\providecommand \@@href[1]{\endgroup#1\@@endlink}%
\providecommand \@sanitize@url [0]{\catcode `\\12\catcode `\$12\catcode
  `\&12\catcode `\#12\catcode `\^12\catcode `\_12\catcode `\%12\relax}%
\providecommand \@@startlink[1]{}%
\providecommand \@@endlink[0]{}%
\providecommand \url  [0]{\begingroup\@sanitize@url \@url }%
\providecommand \@url [1]{\endgroup\@href {#1}{\urlprefix }}%
\providecommand \urlprefix  [0]{URL }%
\providecommand \Eprint [0]{\href }%
\providecommand \doibase [0]{http://dx.doi.org/}%
\providecommand \selectlanguage [0]{\@gobble}%
\providecommand \bibinfo  [0]{\@secondoftwo}%
\providecommand \bibfield  [0]{\@secondoftwo}%
\providecommand \translation [1]{[#1]}%
\providecommand \BibitemOpen [0]{}%
\providecommand \bibitemStop [0]{}%
\providecommand \bibitemNoStop [0]{.\EOS\space}%
\providecommand \EOS [0]{\spacefactor3000\relax}%
\providecommand \BibitemShut  [1]{\csname bibitem#1\endcsname}%
\let\auto@bib@innerbib\@empty
\bibitem [{\citenamefont {Friedan}\ \emph {et~al.}(1985)\citenamefont
  {Friedan}, \citenamefont {Qiu},\ and\ \citenamefont
  {Shenker}}]{Friedan:1984rv}%
  \BibitemOpen
  \bibfield  {author} {\bibinfo {author} {\bibfnamefont {D.}~\bibnamefont
  {Friedan}}, \bibinfo {author} {\bibfnamefont {Z.-a.}\ \bibnamefont {Qiu}}, \
  and\ \bibinfo {author} {\bibfnamefont {S.~H.}\ \bibnamefont {Shenker}},\
  }\href {\doibase 10.1016/0370-2693(85)90819-6} {\bibfield  {journal}
  {\bibinfo  {journal} {Phys. Lett. B}\ }\textbf {\bibinfo {volume} {151}},\
  \bibinfo {pages} {37} (\bibinfo {year} {1985})}\BibitemShut {NoStop}%
\bibitem [{\citenamefont {Lee}(2007)}]{Lee:2006if}%
  \BibitemOpen
  \bibfield  {author} {\bibinfo {author} {\bibfnamefont {S.-S.}\ \bibnamefont
  {Lee}},\ }\href {\doibase 10.1103/PhysRevB.76.075103} {\bibfield  {journal}
  {\bibinfo  {journal} {Phys. Rev. B}\ }\textbf {\bibinfo {volume} {76}},\
  \bibinfo {pages} {075103} (\bibinfo {year} {2007})},\ \Eprint
  {http://arxiv.org/abs/cond-mat/0611658} {arXiv:cond-mat/0611658} \BibitemShut
  {NoStop}%
\bibitem [{\citenamefont {Kaplan}(1984)}]{Kaplan:1983sk}%
  \BibitemOpen
  \bibfield  {author} {\bibinfo {author} {\bibfnamefont {D.~B.}\ \bibnamefont
  {Kaplan}},\ }\href {\doibase 10.1016/0370-2693(84)91172-9} {\bibfield
  {journal} {\bibinfo  {journal} {Phys. Lett. B}\ }\textbf {\bibinfo {volume}
  {136}},\ \bibinfo {pages} {162} (\bibinfo {year} {1984})}\BibitemShut
  {NoStop}%
\bibitem [{\citenamefont {Grover}\ \emph {et~al.}(2014)\citenamefont {Grover},
  \citenamefont {Sheng},\ and\ \citenamefont {Vishwanath}}]{Grover:2013rc}%
  \BibitemOpen
  \bibfield  {author} {\bibinfo {author} {\bibfnamefont {T.}~\bibnamefont
  {Grover}}, \bibinfo {author} {\bibfnamefont {D.~N.}\ \bibnamefont {Sheng}}, \
  and\ \bibinfo {author} {\bibfnamefont {A.}~\bibnamefont {Vishwanath}},\
  }\href {\doibase 10.1126/science.1248253} {\bibfield  {journal} {\bibinfo
  {journal} {Science}\ }\textbf {\bibinfo {volume} {344}},\ \bibinfo {pages}
  {280} (\bibinfo {year} {2014})},\ \Eprint {http://arxiv.org/abs/1301.7449}
  {arXiv:1301.7449 [cond-mat.str-el]} \BibitemShut {NoStop}%
\bibitem [{\citenamefont {Gadde}\ \emph {et~al.}(2015)\citenamefont {Gadde},
  \citenamefont {Razamat},\ and\ \citenamefont {Willett}}]{Gadde:2015xta}%
  \BibitemOpen
  \bibfield  {author} {\bibinfo {author} {\bibfnamefont {A.}~\bibnamefont
  {Gadde}}, \bibinfo {author} {\bibfnamefont {S.~S.}\ \bibnamefont {Razamat}},
  \ and\ \bibinfo {author} {\bibfnamefont {B.}~\bibnamefont {Willett}},\ }\href
  {\doibase 10.1103/PhysRevLett.115.171604} {\bibfield  {journal} {\bibinfo
  {journal} {Phys. Rev. Lett.}\ }\textbf {\bibinfo {volume} {115}},\ \bibinfo
  {pages} {171604} (\bibinfo {year} {2015})},\ \Eprint
  {http://arxiv.org/abs/1505.05834} {arXiv:1505.05834 [hep-th]} \BibitemShut
  {NoStop}%
\bibitem [{\citenamefont {Maruyoshi}\ and\ \citenamefont
  {Song}(2017{\natexlab{a}})}]{Maruyoshi:2016tqk}%
  \BibitemOpen
  \bibfield  {author} {\bibinfo {author} {\bibfnamefont {K.}~\bibnamefont
  {Maruyoshi}}\ and\ \bibinfo {author} {\bibfnamefont {J.}~\bibnamefont
  {Song}},\ }\href {\doibase 10.1103/PhysRevLett.118.151602} {\bibfield
  {journal} {\bibinfo  {journal} {Phys. Rev. Lett.}\ }\textbf {\bibinfo
  {volume} {118}},\ \bibinfo {pages} {151602} (\bibinfo {year}
  {2017}{\natexlab{a}})},\ \Eprint {http://arxiv.org/abs/1606.05632}
  {arXiv:1606.05632 [hep-th]} \BibitemShut {NoStop}%
\bibitem [{\citenamefont {Razamat}\ and\ \citenamefont
  {Zafrir}(2019)}]{Razamat:2019vfd}%
  \BibitemOpen
  \bibfield  {author} {\bibinfo {author} {\bibfnamefont {S.~S.}\ \bibnamefont
  {Razamat}}\ and\ \bibinfo {author} {\bibfnamefont {G.}~\bibnamefont
  {Zafrir}},\ }\href {\doibase 10.1007/JHEP09(2019)046} {\bibfield  {journal}
  {\bibinfo  {journal} {JHEP}\ }\textbf {\bibinfo {volume} {09}},\ \bibinfo
  {pages} {046} (\bibinfo {year} {2019})},\ \Eprint
  {http://arxiv.org/abs/1906.05088} {arXiv:1906.05088 [hep-th]} \BibitemShut
  {NoStop}%
\bibitem [{\citenamefont {Zafrir}(2020)}]{Zafrir:2019hps}%
  \BibitemOpen
  \bibfield  {author} {\bibinfo {author} {\bibfnamefont {G.}~\bibnamefont
  {Zafrir}},\ }\href {\doibase 10.1007/JHEP12(2020)098} {\bibfield  {journal}
  {\bibinfo  {journal} {JHEP}\ }\textbf {\bibinfo {volume} {12}},\ \bibinfo
  {pages} {098} (\bibinfo {year} {2020})},\ \Eprint
  {http://arxiv.org/abs/1912.09348} {arXiv:1912.09348 [hep-th]} \BibitemShut
  {NoStop}%
\bibitem [{\citenamefont {Razamat}\ and\ \citenamefont
  {Zafrir}(2020)}]{Razamat:2020gcc}%
  \BibitemOpen
  \bibfield  {author} {\bibinfo {author} {\bibfnamefont {S.~S.}\ \bibnamefont
  {Razamat}}\ and\ \bibinfo {author} {\bibfnamefont {G.}~\bibnamefont
  {Zafrir}},\ }\href {\doibase 10.1007/JHEP06(2020)176} {\bibfield  {journal}
  {\bibinfo  {journal} {JHEP}\ }\textbf {\bibinfo {volume} {06}},\ \bibinfo
  {pages} {176} (\bibinfo {year} {2020})},\ \Eprint
  {http://arxiv.org/abs/2003.01843} {arXiv:2003.01843 [hep-th]} \BibitemShut
  {NoStop}%
\bibitem [{\citenamefont {Zafrir}(2021)}]{Zafrir:2020epd}%
  \BibitemOpen
  \bibfield  {author} {\bibinfo {author} {\bibfnamefont {G.}~\bibnamefont
  {Zafrir}},\ }\href {\doibase 10.1007/JHEP01(2021)062} {\bibfield  {journal}
  {\bibinfo  {journal} {JHEP}\ }\textbf {\bibinfo {volume} {01}},\ \bibinfo
  {pages} {062} (\bibinfo {year} {2021})},\ \Eprint
  {http://arxiv.org/abs/2007.14955} {arXiv:2007.14955 [hep-th]} \BibitemShut
  {NoStop}%
\bibitem [{\citenamefont {Seiberg}(1995)}]{Seiberg:1994pq}%
  \BibitemOpen
  \bibfield  {author} {\bibinfo {author} {\bibfnamefont {N.}~\bibnamefont
  {Seiberg}},\ }\href {\doibase 10.1016/0550-3213(94)00023-8} {\bibfield
  {journal} {\bibinfo  {journal} {Nucl. Phys. B}\ }\textbf {\bibinfo {volume}
  {435}},\ \bibinfo {pages} {129} (\bibinfo {year} {1995})},\ \Eprint
  {http://arxiv.org/abs/hep-th/9411149} {arXiv:hep-th/9411149} \BibitemShut
  {NoStop}%
\bibitem [{\citenamefont {Leigh}\ and\ \citenamefont
  {Strassler}(1995)}]{Leigh:1995ep}%
  \BibitemOpen
  \bibfield  {author} {\bibinfo {author} {\bibfnamefont {R.~G.}\ \bibnamefont
  {Leigh}}\ and\ \bibinfo {author} {\bibfnamefont {M.~J.}\ \bibnamefont
  {Strassler}},\ }\href {\doibase 10.1016/0550-3213(95)00261-P} {\bibfield
  {journal} {\bibinfo  {journal} {Nucl. Phys. B}\ }\textbf {\bibinfo {volume}
  {447}},\ \bibinfo {pages} {95} (\bibinfo {year} {1995})},\ \Eprint
  {http://arxiv.org/abs/hep-th/9503121} {arXiv:hep-th/9503121} \BibitemShut
  {NoStop}%
\bibitem [{\citenamefont {Argyres}\ and\ \citenamefont
  {Douglas}(1995)}]{Argyres:1995jj}%
  \BibitemOpen
  \bibfield  {author} {\bibinfo {author} {\bibfnamefont {P.~C.}\ \bibnamefont
  {Argyres}}\ and\ \bibinfo {author} {\bibfnamefont {M.~R.}\ \bibnamefont
  {Douglas}},\ }\href {\doibase 10.1016/0550-3213(95)00281-V} {\bibfield
  {journal} {\bibinfo  {journal} {Nucl. Phys. B}\ }\textbf {\bibinfo {volume}
  {448}},\ \bibinfo {pages} {93} (\bibinfo {year} {1995})},\ \Eprint
  {http://arxiv.org/abs/hep-th/9505062} {arXiv:hep-th/9505062} \BibitemShut
  {NoStop}%
\bibitem [{\citenamefont {Xie}(2013)}]{Xie:2012hs}%
  \BibitemOpen
  \bibfield  {author} {\bibinfo {author} {\bibfnamefont {D.}~\bibnamefont
  {Xie}},\ }\href {\doibase 10.1007/JHEP01(2013)100} {\bibfield  {journal}
  {\bibinfo  {journal} {JHEP}\ }\textbf {\bibinfo {volume} {01}},\ \bibinfo
  {pages} {100} (\bibinfo {year} {2013})},\ \Eprint
  {http://arxiv.org/abs/1204.2270} {arXiv:1204.2270 [hep-th]} \BibitemShut
  {NoStop}%
\bibitem [{\citenamefont {Cecotti}\ and\ \citenamefont
  {Del~Zotto}(2013)}]{Cecotti:2012jx}%
  \BibitemOpen
  \bibfield  {author} {\bibinfo {author} {\bibfnamefont {S.}~\bibnamefont
  {Cecotti}}\ and\ \bibinfo {author} {\bibfnamefont {M.}~\bibnamefont
  {Del~Zotto}},\ }\href {\doibase 10.1007/JHEP01(2013)191} {\bibfield
  {journal} {\bibinfo  {journal} {JHEP}\ }\textbf {\bibinfo {volume} {01}},\
  \bibinfo {pages} {191} (\bibinfo {year} {2013})},\ \Eprint
  {http://arxiv.org/abs/1210.2886} {arXiv:1210.2886 [hep-th]} \BibitemShut
  {NoStop}%
\bibitem [{\citenamefont {Cecotti}\ \emph {et~al.}(2013)\citenamefont
  {Cecotti}, \citenamefont {Del~Zotto},\ and\ \citenamefont
  {Giacomelli}}]{Cecotti:2013lda}%
  \BibitemOpen
  \bibfield  {author} {\bibinfo {author} {\bibfnamefont {S.}~\bibnamefont
  {Cecotti}}, \bibinfo {author} {\bibfnamefont {M.}~\bibnamefont {Del~Zotto}},
  \ and\ \bibinfo {author} {\bibfnamefont {S.}~\bibnamefont {Giacomelli}},\
  }\href {\doibase 10.1007/JHEP04(2013)153} {\bibfield  {journal} {\bibinfo
  {journal} {JHEP}\ }\textbf {\bibinfo {volume} {04}},\ \bibinfo {pages} {153}
  (\bibinfo {year} {2013})},\ \Eprint {http://arxiv.org/abs/1303.3149}
  {arXiv:1303.3149 [hep-th]} \BibitemShut {NoStop}%
\bibitem [{\citenamefont {Xie}\ \emph {et~al.}(2021)\citenamefont {Xie},
  \citenamefont {Yan},\ and\ \citenamefont {Yau}}]{Xie:2016evu}%
  \BibitemOpen
  \bibfield  {author} {\bibinfo {author} {\bibfnamefont {D.}~\bibnamefont
  {Xie}}, \bibinfo {author} {\bibfnamefont {W.}~\bibnamefont {Yan}}, \ and\
  \bibinfo {author} {\bibfnamefont {S.-T.}\ \bibnamefont {Yau}},\ }\href
  {\doibase 10.1103/PhysRevD.103.065003} {\bibfield  {journal} {\bibinfo
  {journal} {Phys. Rev. D}\ }\textbf {\bibinfo {volume} {103}},\ \bibinfo
  {pages} {065003} (\bibinfo {year} {2021})},\ \Eprint
  {http://arxiv.org/abs/1604.02155} {arXiv:1604.02155 [hep-th]} \BibitemShut
  {NoStop}%
\bibitem [{\citenamefont {Song}\ \emph {et~al.}(2017)\citenamefont {Song},
  \citenamefont {Xie},\ and\ \citenamefont {Yan}}]{Song:2017oew}%
  \BibitemOpen
  \bibfield  {author} {\bibinfo {author} {\bibfnamefont {J.}~\bibnamefont
  {Song}}, \bibinfo {author} {\bibfnamefont {D.}~\bibnamefont {Xie}}, \ and\
  \bibinfo {author} {\bibfnamefont {W.}~\bibnamefont {Yan}},\ }\href {\doibase
  10.1007/JHEP12(2017)123} {\bibfield  {journal} {\bibinfo  {journal} {JHEP}\
  }\textbf {\bibinfo {volume} {12}},\ \bibinfo {pages} {123} (\bibinfo {year}
  {2017})},\ \Eprint {http://arxiv.org/abs/1706.01607} {arXiv:1706.01607
  [hep-th]} \BibitemShut {NoStop}%
\bibitem [{\citenamefont {Buican}\ and\ \citenamefont
  {Laczko}(2018)}]{Buican:2017rya}%
  \BibitemOpen
  \bibfield  {author} {\bibinfo {author} {\bibfnamefont {M.}~\bibnamefont
  {Buican}}\ and\ \bibinfo {author} {\bibfnamefont {Z.}~\bibnamefont
  {Laczko}},\ }\href {\doibase 10.1103/PhysRevLett.120.081601} {\bibfield
  {journal} {\bibinfo  {journal} {Phys. Rev. Lett.}\ }\textbf {\bibinfo
  {volume} {120}},\ \bibinfo {pages} {081601} (\bibinfo {year} {2018})},\
  \Eprint {http://arxiv.org/abs/1711.09949} {arXiv:1711.09949 [hep-th]}
  \BibitemShut {NoStop}%
\bibitem [{\citenamefont {Intriligator}\ and\ \citenamefont
  {Seiberg}(1995)}]{Intriligator:1995id}%
  \BibitemOpen
  \bibfield  {author} {\bibinfo {author} {\bibfnamefont {K.~A.}\ \bibnamefont
  {Intriligator}}\ and\ \bibinfo {author} {\bibfnamefont {N.}~\bibnamefont
  {Seiberg}},\ }\href {\doibase 10.1016/0550-3213(95)00159-P} {\bibfield
  {journal} {\bibinfo  {journal} {Nucl. Phys. B}\ }\textbf {\bibinfo {volume}
  {444}},\ \bibinfo {pages} {125} (\bibinfo {year} {1995})},\ \Eprint
  {http://arxiv.org/abs/hep-th/9503179} {arXiv:hep-th/9503179} \BibitemShut
  {NoStop}%
\bibitem [{\citenamefont {Kang}\ \emph {et~al.}(2021)\citenamefont {Kang},
  \citenamefont {Lawrie},\ and\ \citenamefont {Song}}]{Kang:2021lic}%
  \BibitemOpen
  \bibfield  {author} {\bibinfo {author} {\bibfnamefont {M.~J.}\ \bibnamefont
  {Kang}}, \bibinfo {author} {\bibfnamefont {C.}~\bibnamefont {Lawrie}}, \ and\
  \bibinfo {author} {\bibfnamefont {J.}~\bibnamefont {Song}},\ }\href {\doibase
  10.1103/PhysRevD.104.105005} {\bibfield  {journal} {\bibinfo  {journal}
  {Phys. Rev. D}\ }\textbf {\bibinfo {volume} {104}},\ \bibinfo {pages}
  {105005} (\bibinfo {year} {2021})},\ \Eprint
  {http://arxiv.org/abs/2106.12579} {arXiv:2106.12579 [hep-th]} \BibitemShut
  {NoStop}%
\bibitem [{\citenamefont {Kang}\ \emph
  {et~al.}(2022{\natexlab{a}})\citenamefont {Kang}, \citenamefont {Lawrie},
  \citenamefont {Lee},\ and\ \citenamefont {Song}}]{Kang:2021ccs}%
  \BibitemOpen
  \bibfield  {author} {\bibinfo {author} {\bibfnamefont {M.~J.}\ \bibnamefont
  {Kang}}, \bibinfo {author} {\bibfnamefont {C.}~\bibnamefont {Lawrie}},
  \bibinfo {author} {\bibfnamefont {K.-H.}\ \bibnamefont {Lee}}, \ and\
  \bibinfo {author} {\bibfnamefont {J.}~\bibnamefont {Song}},\ }\href {\doibase
  10.1103/PhysRevD.105.126006} {\bibfield  {journal} {\bibinfo  {journal}
  {Phys. Rev. D}\ }\textbf {\bibinfo {volume} {105}},\ \bibinfo {pages}
  {126006} (\bibinfo {year} {2022}{\natexlab{a}})},\ \Eprint
  {http://arxiv.org/abs/2111.12092} {arXiv:2111.12092 [hep-th]} \BibitemShut
  {NoStop}%
\bibitem [{\citenamefont {Kang}\ \emph
  {et~al.}(2022{\natexlab{b}})\citenamefont {Kang}, \citenamefont {Lawrie},
  \citenamefont {Lee}, \citenamefont {Sacchi},\ and\ \citenamefont
  {Song}}]{Kang:2022zsl}%
  \BibitemOpen
  \bibfield  {author} {\bibinfo {author} {\bibfnamefont {M.~J.}\ \bibnamefont
  {Kang}}, \bibinfo {author} {\bibfnamefont {C.}~\bibnamefont {Lawrie}},
  \bibinfo {author} {\bibfnamefont {K.-H.}\ \bibnamefont {Lee}}, \bibinfo
  {author} {\bibfnamefont {M.}~\bibnamefont {Sacchi}}, \ and\ \bibinfo {author}
  {\bibfnamefont {J.}~\bibnamefont {Song}},\ }\href {\doibase
  10.1103/PhysRevD.106.106021} {\bibfield  {journal} {\bibinfo  {journal}
  {Phys. Rev. D}\ }\textbf {\bibinfo {volume} {106}},\ \bibinfo {pages}
  {106021} (\bibinfo {year} {2022}{\natexlab{b}})},\ \Eprint
  {http://arxiv.org/abs/2207.05764} {arXiv:2207.05764 [hep-th]} \BibitemShut
  {NoStop}%
\bibitem [{\citenamefont {Kang}\ \emph
  {et~al.}(2023{\natexlab{a}})\citenamefont {Kang}, \citenamefont {Lawrie},
  \citenamefont {Lee},\ and\ \citenamefont {Song}}]{Kang:2022vab}%
  \BibitemOpen
  \bibfield  {author} {\bibinfo {author} {\bibfnamefont {M.~J.}\ \bibnamefont
  {Kang}}, \bibinfo {author} {\bibfnamefont {C.}~\bibnamefont {Lawrie}},
  \bibinfo {author} {\bibfnamefont {K.-H.}\ \bibnamefont {Lee}}, \ and\
  \bibinfo {author} {\bibfnamefont {J.}~\bibnamefont {Song}},\ }\href {\doibase
  10.1103/PhysRevD.107.066018} {\bibfield  {journal} {\bibinfo  {journal}
  {Phys. Rev. D}\ }\textbf {\bibinfo {volume} {107}},\ \bibinfo {pages}
  {066018} (\bibinfo {year} {2023}{\natexlab{a}})},\ \Eprint
  {http://arxiv.org/abs/2210.06497} {arXiv:2210.06497 [hep-th]} \BibitemShut
  {NoStop}%
\bibitem [{\citenamefont {Kang}\ \emph
  {et~al.}(2023{\natexlab{b}})\citenamefont {Kang}, \citenamefont {Lawrie},
  \citenamefont {Lee},\ and\ \citenamefont {Song}}]{LANDSCAPE}%
  \BibitemOpen
  \bibfield  {author} {\bibinfo {author} {\bibfnamefont {M.~J.}\ \bibnamefont
  {Kang}}, \bibinfo {author} {\bibfnamefont {C.}~\bibnamefont {Lawrie}},
  \bibinfo {author} {\bibfnamefont {K.-H.}\ \bibnamefont {Lee}}, \ and\
  \bibinfo {author} {\bibfnamefont {J.}~\bibnamefont {Song}},\ }\href@noop {}
  {\enquote {\bibinfo {title} {{Landscape of 4d SCFTs with $a = c$}},}\
  }\bibinfo {howpublished} {{\it in preparation}} (\bibinfo {year}
  {2023}{\natexlab{b}})\BibitemShut {NoStop}%
\bibitem [{\citenamefont {Gaiotto}(2012)}]{Gaiotto:2009we}%
  \BibitemOpen
  \bibfield  {author} {\bibinfo {author} {\bibfnamefont {D.}~\bibnamefont
  {Gaiotto}},\ }\href {\doibase 10.1007/JHEP08(2012)034} {\bibfield  {journal}
  {\bibinfo  {journal} {JHEP}\ }\textbf {\bibinfo {volume} {08}},\ \bibinfo
  {pages} {034} (\bibinfo {year} {2012})},\ \Eprint
  {http://arxiv.org/abs/0904.2715} {arXiv:0904.2715 [hep-th]} \BibitemShut
  {NoStop}%
\bibitem [{\citenamefont {Gaiotto}\ \emph {et~al.}(2013)\citenamefont
  {Gaiotto}, \citenamefont {Moore},\ and\ \citenamefont
  {Neitzke}}]{Gaiotto:2009hg}%
  \BibitemOpen
  \bibfield  {author} {\bibinfo {author} {\bibfnamefont {D.}~\bibnamefont
  {Gaiotto}}, \bibinfo {author} {\bibfnamefont {G.~W.}\ \bibnamefont {Moore}},
  \ and\ \bibinfo {author} {\bibfnamefont {A.}~\bibnamefont {Neitzke}},\ }\href
  {\doibase 10.1016/j.aim.2012.09.027} {\bibfield  {journal} {\bibinfo
  {journal} {Adv. Math.}\ }\textbf {\bibinfo {volume} {234}},\ \bibinfo {pages}
  {239} (\bibinfo {year} {2013})},\ \Eprint {http://arxiv.org/abs/0907.3987}
  {arXiv:0907.3987 [hep-th]} \BibitemShut {NoStop}%
\bibitem [{\citenamefont {Intriligator}\ and\ \citenamefont
  {Wecht}(2003)}]{Intriligator:2003jj}%
  \BibitemOpen
  \bibfield  {author} {\bibinfo {author} {\bibfnamefont {K.~A.}\ \bibnamefont
  {Intriligator}}\ and\ \bibinfo {author} {\bibfnamefont {B.}~\bibnamefont
  {Wecht}},\ }\href {\doibase 10.1016/S0550-3213(03)00459-0} {\bibfield
  {journal} {\bibinfo  {journal} {Nucl. Phys. B}\ }\textbf {\bibinfo {volume}
  {667}},\ \bibinfo {pages} {183} (\bibinfo {year} {2003})},\ \Eprint
  {http://arxiv.org/abs/hep-th/0304128} {arXiv:hep-th/0304128} \BibitemShut
  {NoStop}%
\bibitem [{\citenamefont {Anselmi}\ \emph {et~al.}(1998)\citenamefont
  {Anselmi}, \citenamefont {Freedman}, \citenamefont {Grisaru},\ and\
  \citenamefont {Johansen}}]{Anselmi:1997am}%
  \BibitemOpen
  \bibfield  {author} {\bibinfo {author} {\bibfnamefont {D.}~\bibnamefont
  {Anselmi}}, \bibinfo {author} {\bibfnamefont {D.~Z.}\ \bibnamefont
  {Freedman}}, \bibinfo {author} {\bibfnamefont {M.~T.}\ \bibnamefont
  {Grisaru}}, \ and\ \bibinfo {author} {\bibfnamefont {A.~A.}\ \bibnamefont
  {Johansen}},\ }\href {\doibase 10.1016/S0550-3213(98)00278-8} {\bibfield
  {journal} {\bibinfo  {journal} {Nucl. Phys. B}\ }\textbf {\bibinfo {volume}
  {526}},\ \bibinfo {pages} {543} (\bibinfo {year} {1998})},\ \Eprint
  {http://arxiv.org/abs/hep-th/9708042} {arXiv:hep-th/9708042} \BibitemShut
  {NoStop}%
\bibitem [{\citenamefont {Agarwal}\ \emph {et~al.}(2019)\citenamefont
  {Agarwal}, \citenamefont {Lee},\ and\ \citenamefont
  {Song}}]{Agarwal:2018zqi}%
  \BibitemOpen
  \bibfield  {author} {\bibinfo {author} {\bibfnamefont {P.}~\bibnamefont
  {Agarwal}}, \bibinfo {author} {\bibfnamefont {S.}~\bibnamefont {Lee}}, \ and\
  \bibinfo {author} {\bibfnamefont {J.}~\bibnamefont {Song}},\ }\href {\doibase
  10.1007/JHEP06(2019)102} {\bibfield  {journal} {\bibinfo  {journal} {JHEP}\
  }\textbf {\bibinfo {volume} {06}},\ \bibinfo {pages} {102} (\bibinfo {year}
  {2019})},\ \Eprint {http://arxiv.org/abs/1812.04743} {arXiv:1812.04743
  [hep-th]} \BibitemShut {NoStop}%
\bibitem [{\citenamefont {Green}\ \emph {et~al.}(2010)\citenamefont {Green},
  \citenamefont {Komargodski}, \citenamefont {Seiberg}, \citenamefont
  {Tachikawa},\ and\ \citenamefont {Wecht}}]{Green:2010da}%
  \BibitemOpen
  \bibfield  {author} {\bibinfo {author} {\bibfnamefont {D.}~\bibnamefont
  {Green}}, \bibinfo {author} {\bibfnamefont {Z.}~\bibnamefont {Komargodski}},
  \bibinfo {author} {\bibfnamefont {N.}~\bibnamefont {Seiberg}}, \bibinfo
  {author} {\bibfnamefont {Y.}~\bibnamefont {Tachikawa}}, \ and\ \bibinfo
  {author} {\bibfnamefont {B.}~\bibnamefont {Wecht}},\ }\href {\doibase
  10.1007/JHEP06(2010)106} {\bibfield  {journal} {\bibinfo  {journal} {JHEP}\
  }\textbf {\bibinfo {volume} {06}},\ \bibinfo {pages} {106} (\bibinfo {year}
  {2010})},\ \Eprint {http://arxiv.org/abs/1005.3546} {arXiv:1005.3546
  [hep-th]} \BibitemShut {NoStop}%
\bibitem [{Note1()}]{Note1}%
  \BibitemOpen
  \bibinfo {note} {There is also a gauge coupling, which breaks a $U(1)$
  symmetry via an ABJ anomaly \cite {Leigh:1995ep}. Therefore the counting of
  the dimension still goes through.}\BibitemShut {Stop}%
\bibitem [{\citenamefont {Kinney}\ \emph {et~al.}(2007)\citenamefont {Kinney},
  \citenamefont {Maldacena}, \citenamefont {Minwalla},\ and\ \citenamefont
  {Raju}}]{Kinney:2005ej}%
  \BibitemOpen
  \bibfield  {author} {\bibinfo {author} {\bibfnamefont {J.}~\bibnamefont
  {Kinney}}, \bibinfo {author} {\bibfnamefont {J.~M.}\ \bibnamefont
  {Maldacena}}, \bibinfo {author} {\bibfnamefont {S.}~\bibnamefont {Minwalla}},
  \ and\ \bibinfo {author} {\bibfnamefont {S.}~\bibnamefont {Raju}},\ }\href
  {\doibase 10.1007/s00220-007-0258-7} {\bibfield  {journal} {\bibinfo
  {journal} {Commun. Math. Phys.}\ }\textbf {\bibinfo {volume} {275}},\
  \bibinfo {pages} {209} (\bibinfo {year} {2007})},\ \Eprint
  {http://arxiv.org/abs/hep-th/0510251} {arXiv:hep-th/0510251} \BibitemShut
  {NoStop}%
\bibitem [{\citenamefont {Romelsberger}(2006)}]{Romelsberger:2005eg}%
  \BibitemOpen
  \bibfield  {author} {\bibinfo {author} {\bibfnamefont {C.}~\bibnamefont
  {Romelsberger}},\ }\href {\doibase 10.1016/j.nuclphysb.2006.03.037}
  {\bibfield  {journal} {\bibinfo  {journal} {Nucl. Phys. B}\ }\textbf
  {\bibinfo {volume} {747}},\ \bibinfo {pages} {329} (\bibinfo {year}
  {2006})},\ \Eprint {http://arxiv.org/abs/hep-th/0510060}
  {arXiv:hep-th/0510060} \BibitemShut {NoStop}%
\bibitem [{\citenamefont {Maruyoshi}\ and\ \citenamefont
  {Song}(2017{\natexlab{b}})}]{Maruyoshi:2016aim}%
  \BibitemOpen
  \bibfield  {author} {\bibinfo {author} {\bibfnamefont {K.}~\bibnamefont
  {Maruyoshi}}\ and\ \bibinfo {author} {\bibfnamefont {J.}~\bibnamefont
  {Song}},\ }\href {\doibase 10.1007/JHEP02(2017)075} {\bibfield  {journal}
  {\bibinfo  {journal} {JHEP}\ }\textbf {\bibinfo {volume} {02}},\ \bibinfo
  {pages} {075} (\bibinfo {year} {2017}{\natexlab{b}})},\ \Eprint
  {http://arxiv.org/abs/1607.04281} {arXiv:1607.04281 [hep-th]} \BibitemShut
  {NoStop}%
\bibitem [{\citenamefont {Agarwal}\ \emph {et~al.}(2016)\citenamefont
  {Agarwal}, \citenamefont {Maruyoshi},\ and\ \citenamefont
  {Song}}]{Agarwal:2016pjo}%
  \BibitemOpen
  \bibfield  {author} {\bibinfo {author} {\bibfnamefont {P.}~\bibnamefont
  {Agarwal}}, \bibinfo {author} {\bibfnamefont {K.}~\bibnamefont {Maruyoshi}},
  \ and\ \bibinfo {author} {\bibfnamefont {J.}~\bibnamefont {Song}},\ }\href
  {\doibase 10.1007/JHEP12(2016)103} {\bibfield  {journal} {\bibinfo  {journal}
  {JHEP}\ }\textbf {\bibinfo {volume} {12}},\ \bibinfo {pages} {103} (\bibinfo
  {year} {2016})},\ \bibinfo {note} {[Addendum: JHEP 04, 113 (2017)]},\ \Eprint
  {http://arxiv.org/abs/1610.05311} {arXiv:1610.05311 [hep-th]} \BibitemShut
  {NoStop}%
\bibitem [{\citenamefont {Gadde}\ \emph {et~al.}(2013)\citenamefont {Gadde},
  \citenamefont {Rastelli}, \citenamefont {Razamat},\ and\ \citenamefont
  {Yan}}]{Gadde:2011uv}%
  \BibitemOpen
  \bibfield  {author} {\bibinfo {author} {\bibfnamefont {A.}~\bibnamefont
  {Gadde}}, \bibinfo {author} {\bibfnamefont {L.}~\bibnamefont {Rastelli}},
  \bibinfo {author} {\bibfnamefont {S.~S.}\ \bibnamefont {Razamat}}, \ and\
  \bibinfo {author} {\bibfnamefont {W.}~\bibnamefont {Yan}},\ }\href {\doibase
  10.1007/s00220-012-1607-8} {\bibfield  {journal} {\bibinfo  {journal}
  {Commun. Math. Phys.}\ }\textbf {\bibinfo {volume} {319}},\ \bibinfo {pages}
  {147} (\bibinfo {year} {2013})},\ \Eprint {http://arxiv.org/abs/1110.3740}
  {arXiv:1110.3740 [hep-th]} \BibitemShut {NoStop}%
\bibitem [{\citenamefont {Buican}\ and\ \citenamefont
  {Nishinaka}(2016)}]{Buican:2016hnq}%
  \BibitemOpen
  \bibfield  {author} {\bibinfo {author} {\bibfnamefont {M.}~\bibnamefont
  {Buican}}\ and\ \bibinfo {author} {\bibfnamefont {T.}~\bibnamefont
  {Nishinaka}},\ }\href {\doibase 10.1103/PhysRevD.94.125002} {\bibfield
  {journal} {\bibinfo  {journal} {Phys. Rev. D}\ }\textbf {\bibinfo {volume}
  {94}},\ \bibinfo {pages} {125002} (\bibinfo {year} {2016})},\ \Eprint
  {http://arxiv.org/abs/1602.05545} {arXiv:1602.05545 [hep-th]} \BibitemShut
  {NoStop}%
\bibitem [{\citenamefont {Bolognesi}\ \emph {et~al.}(2015)\citenamefont
  {Bolognesi}, \citenamefont {Giacomelli},\ and\ \citenamefont
  {Konishi}}]{Bolognesi:2015wta}%
  \BibitemOpen
  \bibfield  {author} {\bibinfo {author} {\bibfnamefont {S.}~\bibnamefont
  {Bolognesi}}, \bibinfo {author} {\bibfnamefont {S.}~\bibnamefont
  {Giacomelli}}, \ and\ \bibinfo {author} {\bibfnamefont {K.}~\bibnamefont
  {Konishi}},\ }\href {\doibase 10.1007/JHEP08(2015)131} {\bibfield  {journal}
  {\bibinfo  {journal} {JHEP}\ }\textbf {\bibinfo {volume} {08}},\ \bibinfo
  {pages} {131} (\bibinfo {year} {2015})},\ \Eprint
  {http://arxiv.org/abs/1505.05801} {arXiv:1505.05801 [hep-th]} \BibitemShut
  {NoStop}%
\bibitem [{\citenamefont {Xie}\ and\ \citenamefont {Yan}(2023)}]{Xie:2021omd}%
  \BibitemOpen
  \bibfield  {author} {\bibinfo {author} {\bibfnamefont {D.}~\bibnamefont
  {Xie}}\ and\ \bibinfo {author} {\bibfnamefont {W.}~\bibnamefont {Yan}},\
  }\href {\doibase 10.1007/JHEP03(2023)201} {\bibfield  {journal} {\bibinfo
  {journal} {JHEP}\ }\textbf {\bibinfo {volume} {03}},\ \bibinfo {pages} {201}
  (\bibinfo {year} {2023})},\ \Eprint {http://arxiv.org/abs/2109.04090}
  {arXiv:2109.04090 [hep-th]} \BibitemShut {NoStop}%
\bibitem [{Note2()}]{Note2}%
  \BibitemOpen
  \bibinfo {note} {When more than three $\protect \mathcal {D}_{p_i}(G)$
  theories are glued, it flows to $\protect \mathcal {N}=4$ SYM only for
  restricted choices of $p_i$; most become IR-free upon linear deformation
  \cite {LANDSCAPE}.}\BibitemShut {Stop}%
\bibitem [{\citenamefont {Bah}\ \emph {et~al.}(2012)\citenamefont {Bah},
  \citenamefont {Beem}, \citenamefont {Bobev},\ and\ \citenamefont
  {Wecht}}]{Bah:2012dg}%
  \BibitemOpen
  \bibfield  {author} {\bibinfo {author} {\bibfnamefont {I.}~\bibnamefont
  {Bah}}, \bibinfo {author} {\bibfnamefont {C.}~\bibnamefont {Beem}}, \bibinfo
  {author} {\bibfnamefont {N.}~\bibnamefont {Bobev}}, \ and\ \bibinfo {author}
  {\bibfnamefont {B.}~\bibnamefont {Wecht}},\ }\href {\doibase
  10.1007/JHEP06(2012)005} {\bibfield  {journal} {\bibinfo  {journal} {JHEP}\
  }\textbf {\bibinfo {volume} {06}},\ \bibinfo {pages} {005} (\bibinfo {year}
  {2012})},\ \Eprint {http://arxiv.org/abs/1203.0303} {arXiv:1203.0303
  [hep-th]} \BibitemShut {NoStop}%
\bibitem [{\citenamefont {Beem}\ and\ \citenamefont
  {Peelaers}(2022)}]{Beem:2020pry}%
  \BibitemOpen
  \bibfield  {author} {\bibinfo {author} {\bibfnamefont {C.}~\bibnamefont
  {Beem}}\ and\ \bibinfo {author} {\bibfnamefont {W.}~\bibnamefont
  {Peelaers}},\ }\href {\doibase 10.21468/SciPostPhys.12.5.172} {\bibfield
  {journal} {\bibinfo  {journal} {SciPost Phys.}\ }\textbf {\bibinfo {volume}
  {12}},\ \bibinfo {pages} {172} (\bibinfo {year} {2022})},\ \Eprint
  {http://arxiv.org/abs/2005.12282} {arXiv:2005.12282 [hep-th]} \BibitemShut
  {NoStop}%
\bibitem [{\citenamefont {Beem}\ and\ \citenamefont
  {Gadde}(2014)}]{Beem:2012yn}%
  \BibitemOpen
  \bibfield  {author} {\bibinfo {author} {\bibfnamefont {C.}~\bibnamefont
  {Beem}}\ and\ \bibinfo {author} {\bibfnamefont {A.}~\bibnamefont {Gadde}},\
  }\href {\doibase 10.1007/JHEP04(2014)036} {\bibfield  {journal} {\bibinfo
  {journal} {JHEP}\ }\textbf {\bibinfo {volume} {04}},\ \bibinfo {pages} {036}
  (\bibinfo {year} {2014})},\ \Eprint {http://arxiv.org/abs/1212.1467}
  {arXiv:1212.1467 [hep-th]} \BibitemShut {NoStop}%
\bibitem [{\citenamefont {Anselmi}\ and\ \citenamefont
  {Kehagias}(1999)}]{Anselmi:1998zb}%
  \BibitemOpen
  \bibfield  {author} {\bibinfo {author} {\bibfnamefont {D.}~\bibnamefont
  {Anselmi}}\ and\ \bibinfo {author} {\bibfnamefont {A.}~\bibnamefont
  {Kehagias}},\ }\href {\doibase 10.1016/S0370-2693(99)00446-3} {\bibfield
  {journal} {\bibinfo  {journal} {Phys. Lett. B}\ }\textbf {\bibinfo {volume}
  {455}},\ \bibinfo {pages} {155} (\bibinfo {year} {1999})},\ \Eprint
  {http://arxiv.org/abs/hep-th/9812092} {arXiv:hep-th/9812092} \BibitemShut
  {NoStop}%
\end{thebibliography}%

\end{document}